# LP-Infinity: overcoming limits to the number of lifetime positions in the COS FUV channel

David J. Sahnow*[a], Christian I. Johnson[a]

[a] Space Telescope Science Institute, 3700 San Martin Drive, Baltimore, MD, USA 21218

## ABSTRACT

Since the Cosmic Origins Spectrograph (COS) was installed on the Hubble Space Telescope (HST) in 2009, thousands of spectra have been collected with the Far Ultraviolet (FUV) channel. Due to the gain sag inherent in the cross delay line detector, regular adjustments to the positions of the spectra have been made, with the seventh and eighth becoming available to Guest Observers in October 2025. Recent improvements to models of the FUV channel, including more extensive optical modeling and better prediction and accommodation of detector gain sag, have allowed us to identify previously excluded locations on the detector for future use, thus requiring additional Lifetime Positions (LPs) to be defined. Since the onboard flight software limited the number of Lifetime Positions to eight, we developed a new method for implementing additional LPs to avoid dealing with the complications of reassigning previously used LPs. This novel strategy, dubbed LP-Infinity, allows an effectively unlimited number of LPs that is now constrained only by the available detector real estate, allowing COS to continue obtaining high-quality FUV spectra well into the 2030s. We discuss the conception, development and implementation of LP-Infinity, its initial use in Cycle 33, and the plans for future LPs starting in 2027.

## 1. INTRODUCTION

The Cosmic Origins Spectrograph (COS) was installed in the Hubble Space Telescope (HST) in 2009, and since that time has obtained spectra of thousands of astronomical objects. Most of the observations have been with the Far Ultraviolet (FUV) channel, which is sensitive from below 900 Å to 2150 Å and has a typical resolving power of ~15,000 with the medium resolution gratings (G130M and G160M) and ~3000 with the low resolution grating (G140L)[1]. This channel uses a photon-counting cross-delay line detector[2] to measure the location of individual photons. Along with the two-dimensional position of each photon event, the photon arrival time and the pulse height (a measure of the gain of the microchannel plates) are recorded. The gain of the MCPs decreases with exposure, leading first to errors in the recorded position of the event. A walk correction module in the CalCOS calibration pipeline is used to correct the position of these low gain events. If the gain falls further, however, it can be low enough that the event can no longer be processed, and the event is lost. Raising the high voltage across the microchannel plates increases the gain, although there are practical and operational limitations on how much gain can be recovered in this way.

Once the gain has sagged too much, spectra are moved to a different position on the detector by adjusting the pointing of the observatory and the position of the COS aperture. Each of these pointing offset positions is known as a Lifetime Position (LP). On orbit, eight LPs have been enabled, with two more planned over the next two years. Since the enabling of LP3 in 2015, multiple LPs have been used simultaneously, with the choice determined by the grating, central wavelength (cenwave), and (sometimes) exposure length. As of July 2026, five LPs are active (Table 1).

## 2. SELECTION OF LIFETIME POSITIONS

A COS Lifetime Position is determined by the position of a target in the HST focal plane in the cross-dispersion direction; as shown in Table 1, they are defined in arcseconds relative to LP1. Offsetting the spectra in this way allows them to fall on different places on the detector (Figure 1). Since the optical system was designed for LP1, resolving power typically decreases as spectra are moved further away from that position. The ideal location for a new LP depends on many factors, including the gain of the detector at the location of the spectra, the resolving power, the amount of overlap with spectra at other LPs, and the locations of detector hot spots and dead spots. In addition, the positions available to each LP is constrained by a variety of factors, including the physical range of the aperture block mechanism, a light leak from the wavelength calibration aperture through the Flat-Field Calibration Aperture, along with presence of increased distortion and dark rates at the edges of the detector (Figure 2).

*sahnow@stsci.edu; phone 1 410 338-6709

Table 1 COS Lifetime Positions

| Lifetime | Enabled | Offset | Usage as of July 2026 |
|---|---|---|---|
| LP1 | 5/11/2009 | 0.0” | Not used |
| LP2 | 7/23/2012 | +3.5” | Not used |
| LP3 | 2/9/2015 | -2.5” | Special G130M observations, G140L |
| LP4 | 10/2/2017 | -5.0” | G140L acquisitions |
| LP5 | 10/4/2021 | +5.4” | G130M/13xx |
| LP6 | 10/3/2022 | +6.5” | Not used |
| LP7 | 11/3/2025 | +8.3” | G140L/1055, 1096, 1222, 1291 |
| LP10 | 11/3/2024 | -3.7” | G160M |
| LP11 | (11/2026) | -6.7” | Planned for 2026 |
| LP12 | (11/2027) | +1.7” | Planned for 2027 |

At the time COS was installed, only one LP was defined. Later the COS Flight Software (FSW) was updated to allow three and then eight LPs. All FUV central wavelengths (cenwaves) started at LP1, and all were moved to LP2 in 2012. Since then, additional LPs have been enabled every few years as the gain sag has increased. Starting with LP3, not all cenwaves have moved to each new LP, and by HST Cycle 32 in 2025, five LPs were being used simultaneously.

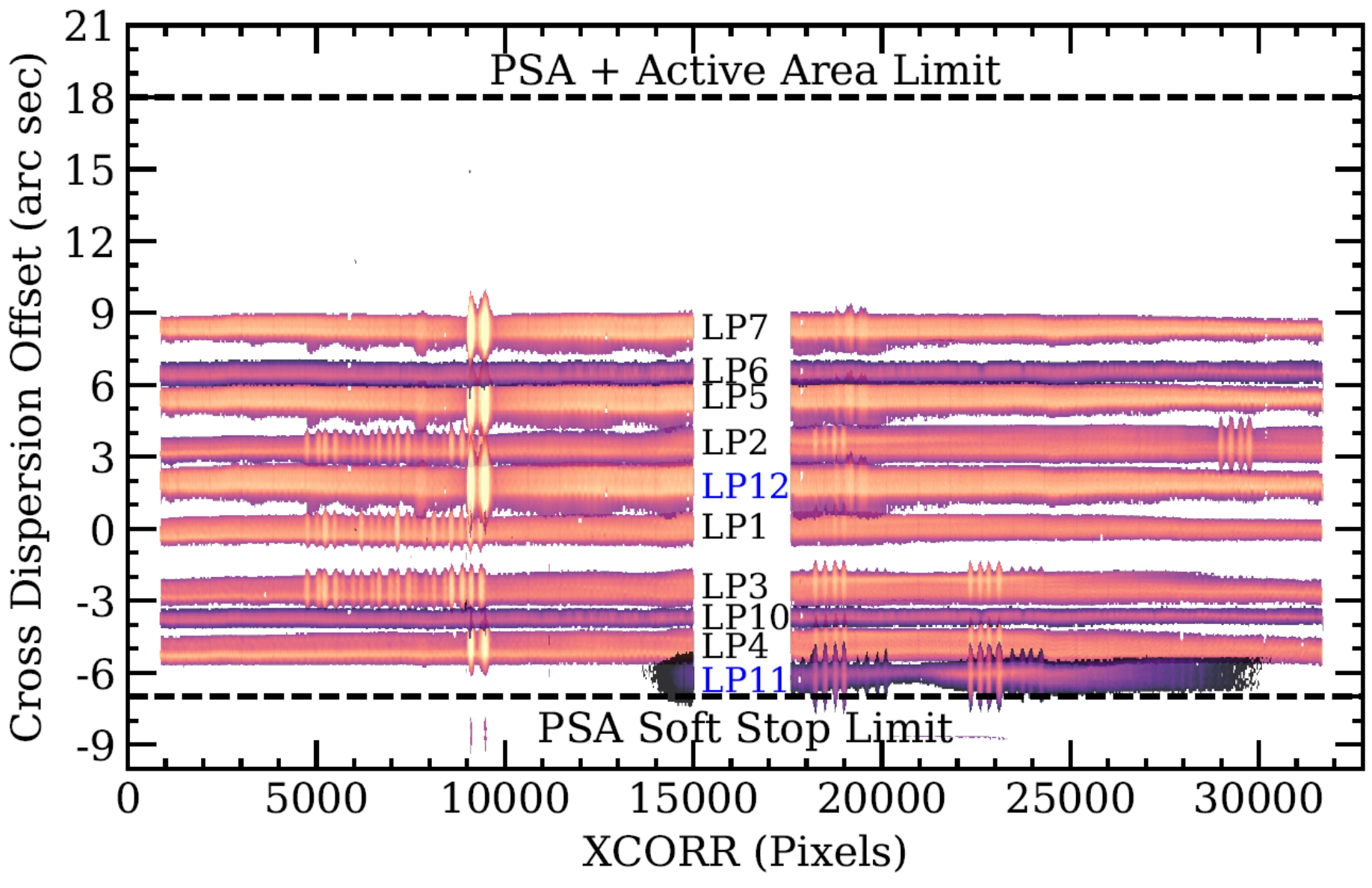

Figure 1 The arrangement of spectra on both detector segments for LP1 – LP12. The extent of each spectrum depends on the combination of central wavelengths that use that LP. The bright, extended features are due to airglow, which fills the aperture, and is seen at either two or four FP-POS positions.

Since the COS apertures are only 2.5 arcseconds in diameter and LPs are typically spaced further apart than that, each requires a unique aperture block position (for each of the two COS apertures). Even at LP1, every central wavelength (cenwave) for each grating had its own focus value. Since the optical path changes slightly at each LP, the best focus value, which is adjusted by changing the position of the OSM1 mechanism that holds the gratings, is different at every LP. These focus positions are also stored in a FSW table.

Although each LP is defined by its position on the sky relative to LP1, the implementation requires the specification of an aperture block position (for each LP) and a grating focus position (for each LP and cenwave). These are stored onboard the instrument in FSW tables.

Each cenwave also has two detector high voltage values (one for each detector segment); these values are adjusted periodically to ensure that the gain on that part of the detector is high enough to ensure good performance. These values are not stored onboard, but are set by Commanding when an observing plan is constructed.

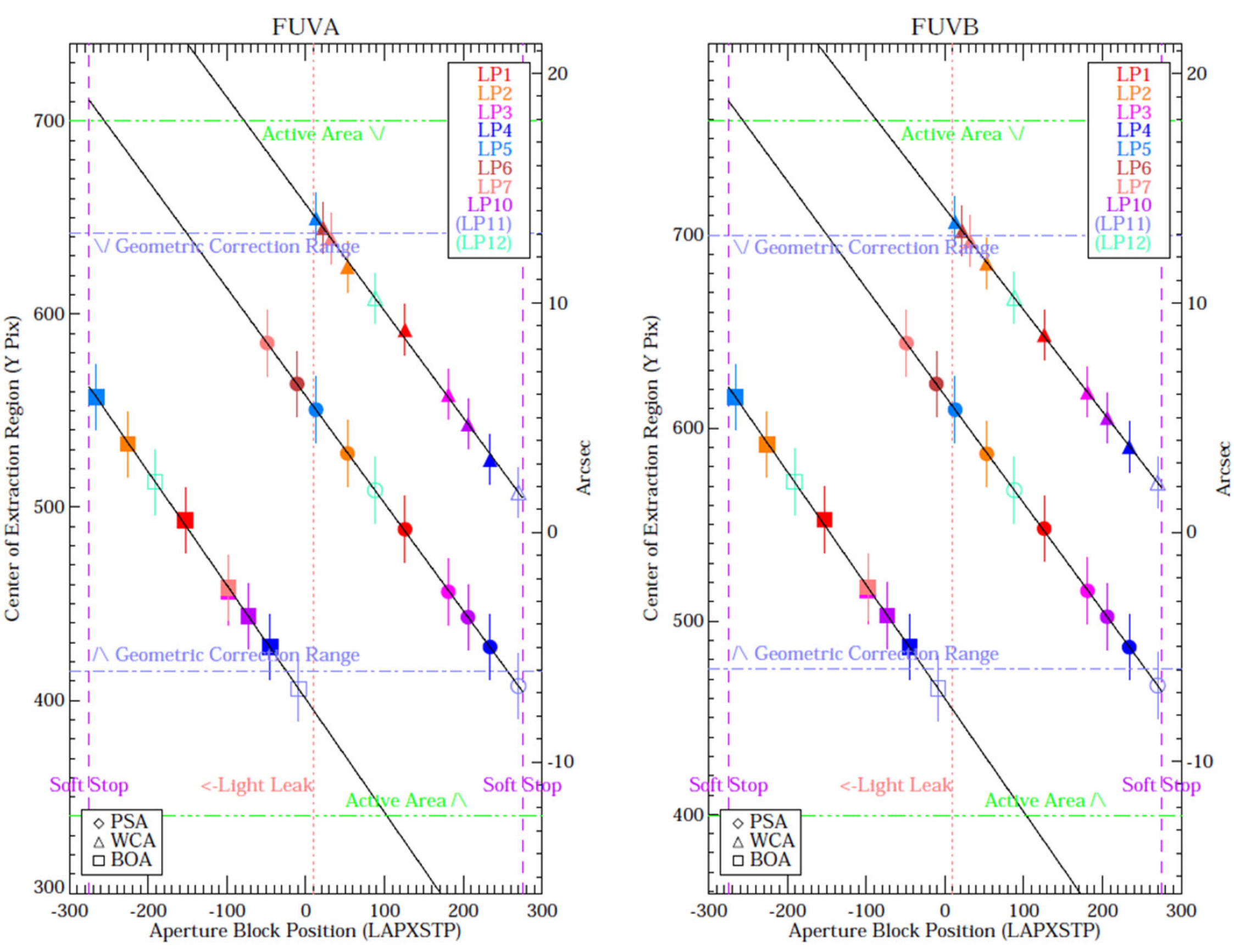

Figure 2 A view of the allocation of the FUV detector real estate showing where the PSA (Primary Science Aperture), BOA (Bright Object Aperture), and WCA (Wavelength Calibration Aperture) fall on the detector for each LP. Limitations due to the aperture block mechanical softs stops, a light leak from the wavelength calibration lamp and increased geometric distortion and dark counts at the top and bottom edges all limit where spectra can be placed.

## 3. LP-INFINITY

The two FSW tables described above currently have room for eight LPs, but as the COS detector aged, it became clear that more than this would be needed in order ensure optimal performance in the long term. The most straightforward solution to this problem would be to increase the size of the tables, but possible issues with onboard memory and the significant resources that are required to modify and test the FSW meant that this would be difficult to accomplish before a ninth LP would be needed. Another option was to reuse the slots in the existing FSW tables by populating previously used entries with new values. However, doing so would have essentially made the LPs time-dependent, and would have complicated both the specification of the exposures and the processing of the data.

After considering the alternatives, an innovative solution that required fewer resources was identified. Instead of increasing the size of the FSW tables, their dimensions could be kept unchanged, but additional LPs could be added by dynamically patching them for each exposure. This methodology, dubbed LP-Infinity (LP∞), meant that the LP8 entries would no longer remain fixed as they had been at previous LPs, but instead, at the beginning of each exposure the table entries corresponding to the desired LP and cenwave would be patched to their desired value, and then the observation would be carried out as if it were an LP8 exposure (Figure 3). This modification required no changes onboard HST, but it did require updates to the Commanding system to handle the patching.

In order to minimize the changes required, LP1 through LP7 were not modified and continue to use the onboard static tables as before. LP8 was skipped to avoid confusion between the old and new methods. To clearly delineate the LP∞ positions, LP9 was also skipped and all new LPs are being assigned two digit numbers, starting with LP10.

Due to faster than expected gain sag on some LPs, the time available to implement LP∞ was limited. Full implementation was completed by the beginning of Cycle 33 in November 2025, only 20 months after the initial concept was developed.

## 4. BENEFITS & COMPLICATIONS

The introduction of the LP-Infinity framework required no change to the pointing methodology, HV commanding, or FSW table size. It simplified testing for all future LPs since no onboard changes are needed when additional LPs are added. It avoided both the reuse of previous LPs and changes to the CalCOS calibration pipeline. Finally, it allows much more flexibility for defining future LPs. For example, multiple LPs at the same detector location can now be defined with different focus values to tailor the highest resolution to a particular wavelength region. This also allows the position of the spectra from each grating (or cenwave) to be more finely controlled, since the optical design, along with slight misalignments between the gratings, leads to different cenwaves projecting to different detector locations at the same LP.

Complications are that the implementation of LP changes for all future LPs moved to the STScI Commanding group and the additional patching required added a few seconds of overhead to each exposure. In addition, there were one-time changes to both APT and TRANS, which are used in the planning and scheduling of observations. Starting in Cycle 34 beginning in late 2026, we have removed default LPs and will require observers to specify the LP directly in their Phase II file. This will simplify the logic that APT maintains as we continue to add LPs. For Cycle 34, only a single LP is permitted for each cenwave, but this may be modified in future cycles. The allowed LPs for each configuration are specified in the COS Instrument Handbook[1] and Phase II Instructions[3].

## 5. COS2035

The use of LP-Infinity, along with previously implemented operational changes that (1) allow non-concurrent wavelength calibration spectra (SPLIT-wavecals); (2) assign different cenwaves to operate at different LPs; (3) use an improved method for flagging gain sagged pixels, and (4) limit the charge extraction in each observing program, should allow COS to operate well into the 2030s with essentially the same performance. Figure 4 shows the current plan for LP allocation going forward based on current estimates of the usage of each cenwave and detector gain sag as a function of exposure. This may be changed going forward, depending on how the detector gain evolves with time. A recent STScI Instrument Science Report provides more details on the plans for extending operations through the 2030s[4].

**Default FSW Table:**

| Cenwave | ... | LP6 | LP7 | LP8 |
|---|---|---|---|---|
| ⋮ | | ⋮ | ⋮ | ⋮ |
| G160M/1533 | ... | -770 | -999 | -999 |
| G160M/1577 | ... | -232 | -384 | -384 |
| ⋮ | | ⋮ | ⋮ | ⋮ |

**G160M/1577 exposure at LP10:**

| Cenwave | ... | LP6 | LP7 | LP8 |
|---|---|---|---|---|
| ⋮ | | ⋮ | ⋮ | ⋮ |
| G160M/1533 | ... | -770 | -999 | -999 |
| G160M/1577 | ... | -232 | -384 | -216 |
| ⋮ | | ⋮ | ⋮ | ⋮ |

**G160M/1533 exposure at LP10:**

| Cenwave | ... | LP6 | LP7 | LP8 |
|---|---|---|---|---|
| ⋮ | | ⋮ | ⋮ | ⋮ |
| G160M/1533 | ... | -770 | -999 | -754 |
| G160M/1577 | ... | -232 | -384 | -216 |
| ⋮ | | ⋮ | ⋮ | ⋮ |

Figure 3 Schematic representation of on-the-fly patching with LP∞ for a portion of the focus FSW Table. The onboard table (top) has valid values for LP6, but default values (blue) for LP7 and LP8. Before a G160M/1577 exposure at LP10 occurs (middle), the LP8 entry is patched with a new value (red). This is followed by a G160M/1533 exposure at LP10 (bottom), which patches a second entry. The values for LP1 through LP7 remain unchanged and continue using the previous method.

COS is already more than a dozen years beyond its 5-year design lifetime, and these innovations will help ensure that the FUV detector is not the limiting factor for obtaining high-quality FUV spectra for as long as possible between now and when the Habitable Worlds Observatory is operating.

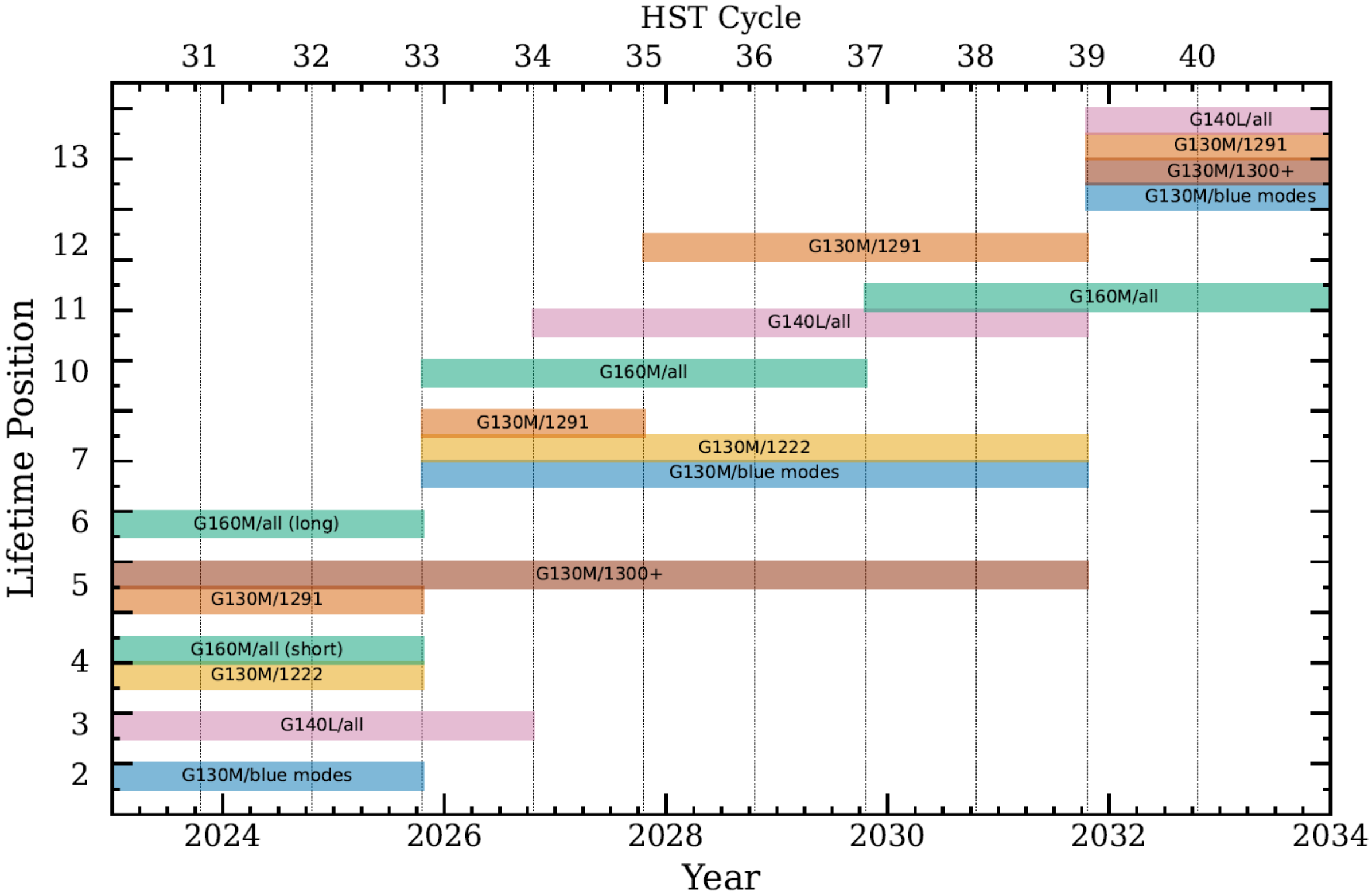

Figure 4 Current plans for LP assignments into the 2030s. These may change depending on detector usage and gain sag. LP1 – LP7 use the original LP method, while two-digit LPs use LP∞.

## ACKNOWLEDGEMENTS

The implementation of LP∞ would not have been possible without the contributions from a large number of colleagues. This includes the COS Branch at STScI, along with members of the Commanding, APT, and TRANS groups at STScI, and our partners at the Goddard Space Flight Center.

## REFERENCES

[1] Payne, A.V. and Dixon, W.V., Cosmic Origins Spectrograph Instrument Handbook, Version 18.0, (2026), https://hst-docs.stsci.edu/cosihb.

[2] McPhate, J. B., Vallerga, J. V., Siegmund, O. H. W., Sahnow, D. J., Penton, S. V., Ake, T. B., France, K., Massa, D., Osterman, S. N., Béland, S., York, B. R. and Welty, A., "HST-COS FUV detector initial on-orbit performance," Proc. SPIE 7732, 7732H (2010).

[3] https://hst-docs.stsci.edu/hpiom

[4] Rafelski, M. et al., "COS2035: Extending COS/FUV Operations Through the 2030s," COS ISR 206-TBD (2026).